\documentclass[twocolumn,showpacs,pra,aps]{revtex4}
\usepackage{amssymb}
\usepackage{amsmath}
\usepackage{amsfonts}
\usepackage{graphics}
\usepackage{epic}
\usepackage{eepic}
\usepackage{color}
\usepackage{epsfig}
\usepackage{bbm}
\usepackage{graphicx}

\begin{document}

\def\ra{\rangle}
\def\la{\langle}
\def\bege{\begin{equation}
  }

\def\ende{\end{equation}}

\def\begarr{\begin{eqnarray}
  }

\def\endarr{\end{eqnarray}}
\def\ha{{\hat a}}
\def\hb{{\hat b}}
\def\hu{{\hat u}}
\def\hv{{\hat v}}
\def\hc{{\hat c}}
\def\hd{{\hat d}}
\def\no{\noindent}\def\non{\nonumber}
\def\hi{\hangindent=45pt}
\def\v{\vskip 12pt}

\newcommand{\bra}[1]{\left\langle #1 \right\vert}
\newcommand{\ket}[1]{\left\vert #1 \right\rangle}

\title{Geometric Phase Based Quantum Computation Applied to an
NP-Complete Problem}

\author{David R.\ Mitchell\footnote{E--mail: mitchellda@nsula.edu }}
\affiliation{Northwestern State University, Natchitoches,
Louisiana 71497\\ }

\date{\today}


\begin{abstract}
We present a new approach to quantum computation involving the
geometric phase.  In this approach, an entire computation is
performed by adiabatically evolving a suitably chosen quantum
system in a closed circuit in parameter space.  The problem solved
is the determination of the solubility of a 3-SAT Boolean
Satisfiability problem.  The problem of non-adiabatic transitions
to higher levels is addressed in several ways.  The avoided level
crossings are well defined and the interpolation can be slowed in
this region, the Hamiltonian can be scaled with problem dimension
resulting in a constant gap size and location, and the
prescription here is sufficiently general as to allow for other
suitably chosen Hamiltonians. Finally, we show that with $n$
applications of this approach, the geometric phase based quantum
computation method may be used to find the solution to the 3-SAT
problem in $n$ variables, a member of the NP-complete complexity
class.
\end{abstract}

\maketitle

\section{Introduction}
Geometric phases have been proposed in the context of quantum
computation, specifically in the proposal of fault-tolerant
gates\cite{Jones}\cite{Falce}\cite{Duam}. Here we discuss the
possibility of performing an entire computation using a single
application of the geometric phase.  This approach uses the
geometric phase phenomenon in conjunction with the adiabatic
approach of Farhi, et. al. \cite{Farhi}.  With a single
application, the problem solved is determining the solubility of a
Boolean satisfiability 3-SAT problem in $n$ variables. With $n$
applications of this approach, the geometric phase based quantum
computation method may be used to find the solution to the 3-SAT
problem itself.  This last problem is of the NP-complete
complexity class\cite{Nielson}.

\section{Adiabatic Quantum Computation}
Farhi and coworkers\cite{Farhi} presented an approach to quantum
computation, Adiabatic Quantum Computation (AQC).  The AQC
approach involves an adiabatic interpolation from the ground state
of an initial Hamiltonian ($H_i$) to the ground state of a final
Hamiltonian ($H_p$) that encodes the solution to the problem of
interest.  The method involves adiabatically evolving the
Hamiltonian $H(z) = z H_i + (1-z) H_p$, while remaining in the
ground state. In this approach, the problem solved is an
NP-Complete problem, the 3-SAT Boolean Satisfiability problem
containing m clauses in $n$ variables.  In our construction, each
eigenstate of the $N=2^n$ dimensional $H_p$ encodes a possible
solution to the problem.  The energy of each level of $H_p$
corresponds to the number of clauses that have been violated by
that possible solution, with the ground state E=0 level
corresponding to a solution of the problem (0 violated clauses).

In the context of AQC, it was shown that in the case of a
computationally hard instance of the 3-SAT problem, there exists a
parameter region in which the spectrum is irregular, the spectral
fluctuation distribution taking the form of the Wigner
distribution\cite{Mitchell}. In these regions, it was shown the
interpolation was susceptible to non-adiabatic transitions from
the ground state leading to an interpolation time that scales
exponentially with problem size. The question arises whether
similar AQC ideas could be implemented without interpolating
through the irregular spectral region. Namely, by adding a second
parameter dependence and interpolating around an appropriately
modified final Hamiltonian, the geometric phase may be used and a
computation performed without traversing through a problematic
region.

\section{Geometric Phase}

Berry\cite{Berry} noted that an eigenstate having a parameter
dependence acquires, in addition to the familiar dynamical phase,
a geometric phase (Berry phase) when the state is adiabatically
evolved in a cyclic manner near a level degeneracy, termed a
diabolical point. In the special case of a real Hamiltonian,
cyclic evolution of the parameters produces a phase of $\pi$, an
overall sign change in the eigenstate being transported.

In the current approach, we use a construction similar to that of
AQC - although instead of interpolating in one parameter (z) from
$H_i$ ($z=1$) to $H_p$ ($z=0$), we use two parameters (x, z) and
evolve the system in a closed circuit in the two dimensional
parameter space about the point (0,0).  To illustrate the method,
we show how the Berry phase phenomena yields a test for solubility
of the 3-SAT using $H_p$ itself.  $H_p$ encodes a particular 3-SAT
problem instance in $n$ variables and is constructed easily
without a priori knowledge of the solution\cite{Farhietal2000}.

The Hamiltonian for this problem is constructed in such a way that
a level degeneracy only exists at the point (0,0) if a solution
exists, and no level degeneracy exists if a solution does not
exist.  When the system is transported along the closed path
encircling the point (0,0), the state accrues a Berry phase if a
solution exists and accrues no phase if a solution does not exist.
Since the Hamiltonian is real, the accrued phase becomes $\pi$ in
the event of the existence of a solution\cite{Berry}.

\section{Geometric Phase Based Quantum Computation}
The Hamiltonian we use for this problem is the direct sum of two
Hamiltonians with the addition of interaction terms. The first
Hamiltonian, $H_a$, is constructed by adding
$\frac{z}{4}\mathbb{I}$ to the $H_p$ above: $H_a = H_p +
\frac{z}{4}\mathbb{I}$, which has $N=2^n$ dimensions, where $n$ is
the number of variables in the 3-SAT problem instance. We
construct $H_a$ in such a way that relies on no a priori knowledge
of which state of $H_p$ is the solution state.  This definition of
$H_a$ is selected for certain simplifications, although many forms
for $H_a$ may be chosen.  In fact, it is worth noting that $H_a$
can be chosen to be $H_p$ identically. In the following, we
demonstrate the method for a soluble 3-SAT instance, with the
generalization to non-soluble cases being straightforward.  We
arrange the basis states of $H_p$ such that $H_{p_{ii}} \neq 0$, $
1 \leq i \leq N-1$; and $H_{p_{NN}} = 0$ (the solution state). The
second Hamiltonian under consideration, in addition to $H_a$, is
the Hamiltonian $H_b = -\frac{z}{4}$, having dimension 1.  We
define the unperturbed Hamiltonian for the problem to be $H^{(0)}
= H_a \bigoplus H_b$. The unperturbed states
$\ket{\psi_i}$,$\ket{\psi_a}$, and $\ket{\psi_b}$ are defined by:
\begin{eqnarray}
 H^{(0)} \mid \psi_i \rangle & = & H^{(0)}_{ii} \mid
 \psi_i \rangle = (\frac{z}{4} + H_{p_{ii}})\mid
 \psi_i \rangle   \; ,  (i < N) \nonumber \\
 H^{(0)} \mid \psi_a \rangle & = & H^{(0)}_{NN} \mid \psi_a \rangle
 = \frac{z}{4}\mid \psi_a \rangle  \nonumber \\
 H^{(0)} \mid \psi_b \rangle & = & H^{(0)}_{N+1,N+1} \mid \psi_b \rangle
 = -\frac{z}{4}\mid \psi_b \rangle  \; .  \label{unperturbed}
\end{eqnarray}
The final form for the Hamiltonian used in the geometric phase
computation includes coupling between $H_b$ and all states of
$H_a$ through the additional parameter x.  In the unperturbed
basis of Eq.(\ref{unperturbed}), the general matrix representation
of H that determines the 3-SAT solubility through the geometric
phase phenomenon is
\begin{eqnarray}
H_{ii} = Ha_{ii} = \frac{z}{4} + H_{p_{ii}} \;  & & (i<N) \nonumber \\
H_{NN} = Ha_{NN} = \frac{z}{4} + H_{p_{NN}}  & &  \nonumber \\
H_{iN} = H_{Ni} = x  \;   & & (i<N) \nonumber \\
H_{N+1,N+1}  =  H_b = -\frac{z}{4} & &   \; , \label{unscaled}
\end{eqnarray}
where no particular problem instance has yet been chosen for
$H_p$.

The problem to be solved is determining the existence of a
solution to the 3-SAT problem, or 3-SAT solubility.  Evolving the
ground state of Eq.(\ref{unscaled}) adiabatically in a closed
circuit in the parameters (x,z) results in one of two possible
outcomes: no geometric phase in the case of no solution, or a
geometric phase = $\pi$ in the case of the existence of a solution
to the 3-SAT problem.
\begin{figure}[t]
\includegraphics[width=9cm]{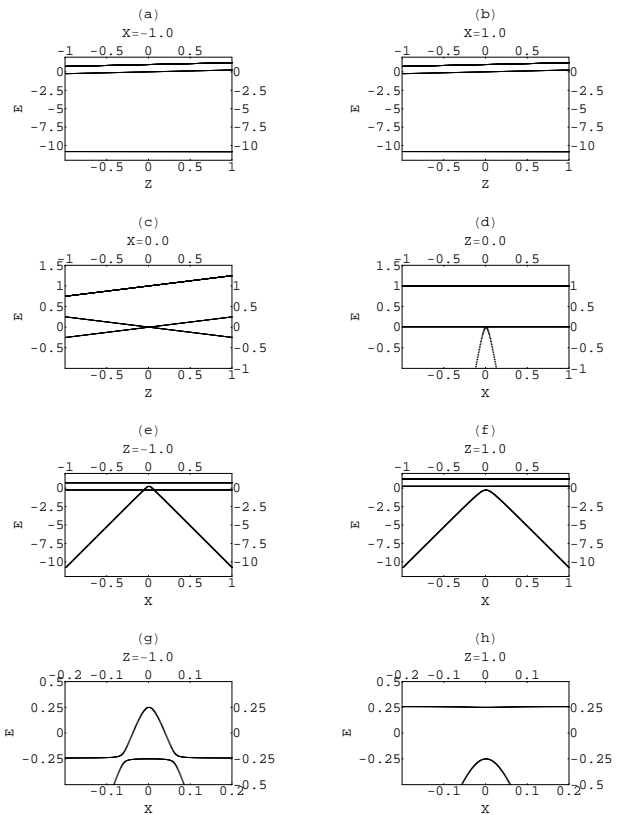}\\
\bigskip
\caption{Spectrum generated by Eq. (1) for varying z: (a),(b);
varying x: (e),(f); the level crossing: (c),(d); and avoided level
crossings: (g). Note: scale varies between rows.  Note also
(g),(h) are identical to (e),(f) with a change of scale. The path
in parameter space traverses frames (f) (r to l) $ \rightarrow $
(a) (r to l) $\rightarrow $ (e) (l to r) $ \rightarrow $ (b) (l to
r) $ \rightarrow $ (f).\label{fig1}}
\end{figure}

Our numerical analysis uses the specific case of n=7, N=128. In
addition, we choose a specific problem instance encoded by $H_p$.
We choose the highly degenerate case of $H_p$ having a solution
state (E=0) and N-1 degenerate states (E=1). Specifically,
$H_{p_{ii}}=1$, $i<N$ and $H_{p_{NN}}=0$. Although somewhat
unrealistic from a problem standpoint, this case produces the
smallest minimum gap size at the avoided level crossings (see Fig.
1g) where a non-adiabatic transition may occur, and so represents
the worst case situation. This is discussed further below.

We start the system in the ground state of the system at
(x,z)=(0,1) (see Fig. 1h); this corresponds to the unperturbed
state $\ket{\psi_b}$, with the eigenvalue equation $H(0,1)
\ket{\psi_b(0,1)} = -\frac{1}{4} \ket{\psi_b(0,1)} $. This state
is then adiabatically evolved around the following parametric
circuit: $(x,z)=(0,1)\rightarrow (-1,1)\rightarrow(-1,-1)
\rightarrow(1,-1)\rightarrow (1,1)\rightarrow (0,1)$.

The approach may be viewed graphically in the following manner.
The spectrum generated by Eq.(\ref{unscaled}) with the particular
instance of $H_p$ so defined is shown in Fig. 1.  This choice of
$H_p$ produces a spectrum of three energies levels. The highest
energy sheet has a degeneracy (N-1), and the two lowest are
nondegenerate. These lowest two energy sheets are responsible for
the generation of the geometric phase.

If the 3-SAT problem encoded in $H_p$ has a solution, then the
ground state energy of $H_p$ is 0 and there will be a degeneracy
in the two lowest energy sheets at the origin in parameter space,
($x=0, z=0$)(see Figs. 1c, 1d). When the ground state is evolved
adiabatically in a closed circuit in parameter space about this
point, this will give a Berry phase $= \pi $ (see Figs. 1a, 1b,
1e, 1f).  If $H_p$ is chosen such that no solution exists, there
is no degeneracy, and the closed circuit does not encircle a
diabolical point resulting in no Berry phase.

\section{Avoided Level Crossings}

We now consider the problem of avoided level crossings: gap size
and location.

In order to encircle the origin and accrue the Berry phase, the
state must remain on the ground state energy sheet and not make a
transition at the avoided level crossing (eg. by way of
Landau-Zener transition \cite{Zener})(see Fig. 1g).  Simulations
have shown that the avoided level crossing gap size produced by
Eq.(\ref{unscaled}) with $H_p$ defined above decreases
exponentially with increasing problem size. This is similar to
difficulties in the AQC approach \cite{Mitchell}.  Even so, unlike
other approaches, the location of the avoided level crossings is
well defined, can be predicted to some extent, and the
interpolation slowed in that region to prevent a level transition.

We gain insight into the Hamiltonian in Eq.(\ref{unscaled}) by
calculating the approximate location of the avoided level
crossings using perturbation theory.  We consider the Hamiltonian
in Eq.(\ref{unscaled}) with $H_p$ defined above evaluated at z=-1
and x small (corresponding to Fig. 1g). In this case, the
unperturbed states are given by Eq.(\ref{unperturbed}), taking x
as the perturbation parameter. As x does not appear as a diagonal
entry in the perturbation matrix, there is no first order
correction. The second order correction is easily evaluated using
the following unperturbed energies:
\begin{eqnarray}
E_i^{(0)} = H^{(0)}_{ii} = H_{ii} = 1 + \frac{z}{4} & & (i<N) \nonumber \\
E_a^{(0)} = H^{(0)}_{NN} = H_{NN} = \frac{z}{4} \nonumber \\
E_b^{(0)} = H_{N+1,N+1} = -\frac{z}{4} \label{E0}
\end{eqnarray}
We are interested in the second order perturbative corrections to
$E_a^{(0)}$ and $E_b^{(0)}$, and we denote these energies by
$\Delta_a^{(2)}$ and $\Delta_b^{(2)}$. The second order correction
is given by the familiar formula \cite{Schiff1955} $\Delta_m^{(2)}
= \sum_{n \neq m} \frac{\mid H^\prime_{mn} \mid ^2}{E_m^{(0)} -
E_n^{(0)}}$, where $H^\prime$ is the perturbation matrix
containing only off diagonal terms equal to x ($ H^\prime_{i,N+1}
= H^\prime_{N+1,i} = x, i \leq N $). We begin with
$\Delta_b^{(2)}$. The sum in this case can be split into two
terms, one involving the (N-1) states with energy $E_i^{(0)}$ and
one involving the state with energy $E_a^{(0)}$. In both cases the
matrix element yields $\mid H^\prime_{b,n} \mid ^2 = x^2 (n \leq
N)$. $\Delta_a^{(2)}$ simplifies directly since $H^\prime_{a,i}=0$
for $i \neq N+1$, and the sum reduces to one term.  Evalution of
these sums, using Eq.(\ref{E0}), yields
\begin{eqnarray}
\Delta_a^{(2)} & = & -2x^2 \nonumber \\
\Delta_b^{(2)} & = & \frac{-2(N-1)x^2}{2+z} - \frac{2x^2}{z} \; .
\label{D2}
\end{eqnarray}
Inserting N=128, z=-1 into Eqs.(\ref{E0}),(\ref{D2}) the energies
to 2nd order are:
\begin{eqnarray}
E_a & = & -\frac{1}{4} - 2x^2 \nonumber \\
E_b & = & \frac{1}{4} - 252x^2 \; . \label{Energies}
\end{eqnarray}
This is consistent with Fig. 1e and Fig. 1g.  The differences are
attributed to higher order perturbations,which become more
pronounced with increasing x. The intersection of these two
parabolas gives the approximate location of the avoided level
crossings to second order, $x_{gap}\approx \pm 0.045$ (see Fig.
1g).

Earlier we chose the Hamiltonian $H_{p}$ such that $H_{p_{ii}}=1$,
$i<N$ and $H_{p_{NN}}=0$.  The reason results from the
perturbation theory calculation above:  each level above 0
contributes to the level repulsion on state $\ket{\psi_b}$ (see
Fig. 1g). In this case for $ \mid x \mid  < x_{gap}$, (N-1) levels
contribute to the level repulsion downward, and one state
($\ket{\psi_a}$) repels the state upward.  In the general case,
the sum is dominated by the closest levels.  The $H_p$ chosen has
N-1 levels located at the closest energy corresponding to a
non-solution (E=1).  This results in the situation having the
strongest level repulsion, and in this sense represents the worst
case situation in terms of the avoided level crossings.  Other
choices of 3-SAT problem instances for $H_p$ result in weaker
level repulsion.

\section{Scaling}

In addition to varying the interpolation speed near the avoided
level crossings, the problem can be scaled with N such that the
minimum gap size remains N independent for all problem sizes.  We
choose the scaling such that $\Delta_b^{(2)}$ is N independent for
large N in Eq.(\ref{D2}).  This may be accomplished by scaling z
or by scaling x appropriately.  If z is chosen to scale with N
such that $z \rightarrow Nz$, the pertubative corrections
$\Delta_a^{(2)}$ and $\Delta_b^{(2)}$ approach a constant for
large N.  Explicitly, the Hamiltonian containing this scaling is
the following (cf. Eq.(\ref{unscaled})):
\begin{eqnarray}
H_{ii} = Ha_{ii} = z/4+N H_{p_{ii}} & & 1<i<N \nonumber \\
H_{NN} = Ha_{NN} = z/4+N H_{p_{NN}}   \nonumber \\
H_{iN} = H_{Ni} = x & & 1<i<N \nonumber \\
H_{kk} = H_b = -z/4 & & k=N+1  \; . \label{scaled}
\end{eqnarray}
We have used n=7, N=128 in the numerical simulation of Eq.
(\ref{scaled}), with the results shown in Fig. 2 (cf. Fig. 1). The
scaling was checked numerically for $3 \leq n \leq 9$.  For these
values, a constant minimum gap size ($\Delta_{gap} \approx 0.5$)
and constant gap location ($x_{gap} \approx 0.0 $) was found.
\begin{figure}[t]
\includegraphics[width=9cm]{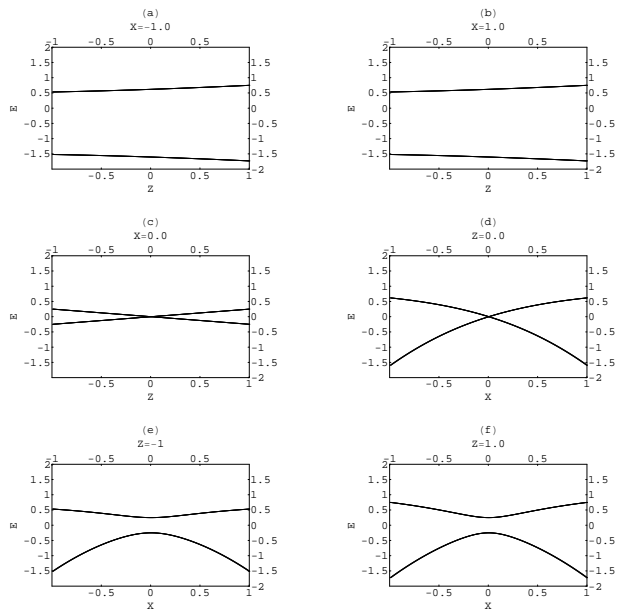}\\
\bigskip
\caption{Energy level plots showing the crossing and avoided level
crossing along the path in parameter space including the scaling
in Eq. (\ref{scaled}). Not shown are the (N-1) degenerate levels
at Energy=$z/4+N H_{p_{ii}} \approx 128 $. \label{fig1}}
\end{figure}

Analogously, we may choose N scaling for x, such that
$\Delta_a^{(2)}$ and $\Delta_b^{(2)}$ is approximately constant
for large N.  This may be accomplished by choosing $x^2$ to scale
with $\frac{1}{N}$, or $x \rightarrow \frac{x}{\sqrt{N}}$.

\section{Solving the 3-SAT problem}
The geometric phase based approach above determines the existence
of a solution of the encoded 3-SAT problem in $n$ variables.  We
now show that upon repeated iterations of this approach, the
solution to the 3-SAT problem may be searched for and the solution
found in $n$ iterations.

As described above, if a solution state exists a geometric phase
will be generated. Initially, each eigenvector $\ket{\phi_i}$ of
$H_a$ encodes a possible solution to the 3-SAT problem.  The
collection of these eigenvectors forms the basis of $H_a$ and
encodes all possible solutions.   This represents the search space
under consideration. From among this set we denote the state
$\ket{\phi_a}$ as the solution state, if it exists.

Using this $H_a$, we construct $H$ as in Eq.(\ref{unscaled}). If
the Berry phase is generated, then the solution state exists in
the search space.  If no phase is generated, then no solution
exists and the search is complete.

In the case of the existence of a solution, we then divide the
search space in half, such that $H_a = H_\alpha^{(1)} \bigoplus
H_\beta^{(1)}$, with the basis of each subspace forming a solution
space, each containing $\frac{N}{2}$ solutions.

Substituting $H_\alpha^{(1)}$ for $H_p$ in Eq.(\ref{unscaled}), we
construct a new Hamiltonian, $H^{(1)}$, having $\frac{N}{2}+1$
dimensions. Performing the geometric phase approach with this
$H^{(1)}$, the system either accrues a phase, in which case the
solution state exists in the basis of $H_\alpha^{(1)}$; or the
system accrues no phase, implying the solution lies in the basis
of $H_\beta^{(1)}$.

The Hamiltonian containing the solution state in its basis is then
subdivided and used in the next iteration in a similar manner. For
example, if $H_\alpha^{(1)}$ generates a phase, then the solution
exists in the basis of $H_\alpha^{(1)}$ which is then subdivided
into $H_\alpha^{(1)} = H_\alpha^{(2)} \bigoplus H_\beta^{(2)}$.
Substituting $H_\alpha^{(2)}$ for $H_p$ in Eq.(\ref{unscaled}), we
construct a new Hamiltonian, $H^{(2)}$, having $\frac{N}{4}+1$
dimensions. A single application of the geometric phase approach
determines the solution to be in the basis of $H_\alpha^{(2)}$ or
$H_\beta^{(2)}$.

This iterative procedure is a sorting algorithm which halves the
search space at each iteration, and locates the half containing
the solution in a single evaluation (the geometric phase based
procedure). Starting with N states, the search is completed in
$\log_2 (N)$ evaluations. In the case of the 3-SAT problem in $n$
variables, the initial search space is $N=2^n$ yielding the
solution in $n$ evaluations.

\section{Conclusion}
We have illustrated a general approach to solving NP-complete
problems using the geometric phase phenomenon in conjunction with
the Adiabatic Quantum Computation method.

We first address the problem of whether a Boolean Satisfiability
problem (3-SAT) is soluble - if so, the adiabatically evolved
eigenstate acquires a geometric phase; if the problem is not
soluble, the state does not acquire a phase.

The problem of avoided level crossings having small gap sizes is
addressed in several ways.  First, the crossings are isolated and
well defined, and so the interpolation can be slowed in this
vicinity.  Second, the Hamiltonian can be scaled with N (see
Eq.(\ref{scaled})) to make the gap sizes independent of problem
size. Third, as mentioned earlier, the Hamiltonian
(Eq.(\ref{unscaled})) is not the only choice which leads to a
successful approach.  This is a particularly simple construction,
but there is a large degree of freedom in choosing the Hamiltonian
that governs the approach.  Other Hamiltonians can be found that
also adhere to additional experimental requirements or parameters
of a different problem.

Repeated iteration of this procedure results in a search of the
set of N basis states encoding the possible solutions that
completes in $\log_2(N)$ iterations. For a problem instance having
$n$ variables, the solution of the 3-SAT can be found in $n$
iterations of the geometric phase based approach.  In this way, we
show the geometric phased based approach yields the solution to
the 3-SAT, an NP-complete problem.

\section{Acknowledgements}

This work was initiated in the Quantum Computation Technology
group at the Jet Propulsion Laboratory while supported by the NASA
Faculty Fellowship Program (NFFP). We would like to thank Jonathan
Dowling for encouragement and helpful comments.

\end{document}